\documentstyle[twoside,fleqn,espcrc2mw,epsf,bm]{article}

\newcommand{\VEC}[1]{{\bf \bm{#1}}} % for bold 3-vectors

\title{ 
Exploring Lattice Methods for Cold Fermionic Atoms
} % end title

\author{
Matthew Wingate\address{Institute for Nuclear Theory,
University of Washington, Seattle, WA 98195-1550}
}

\begin{document}

\begin{abstract}
There has been a surge of experimental effort recently in 
cooling trapped fermionic atoms to quantum degeneracy.  
By varying an external magnetic field, interactions between 
atoms can be made arbitrarily strong.   When the S wave scattering 
length becomes comparable to and larger than the interparticle 
spacing, standard mean field analysis breaks down.   In this case
the system exhibits a type of universality, and J-W.~Chen
and D.B.~Kaplan \cite{Chen:2003vy} recently showed how this system can 
be studied from first principles using lattice field theory.  
This poster presents the first results of exploratory simulations.
The existence of a continuum limit is checked
and the pairing condensate is studied as a function of the external 
source strength over a range of temperatures.  Preliminary results
show simulations can locate the critical temperature.
\end{abstract}

\maketitle

%%%%%
\section{INTRODUCTION}

At $T=0$ the physics of a dilute gas 
is characterized by just two length scales:
the interparticle spacing $n^{-1/3} \equiv \left(\frac{V}{N}\right)^{1/3}$
and the S wave scattering length $a$.
The range of interactions is very small compared
to the average interparticle spacing $R \ll n^{-1/3}$, 
so only 2-body S wave scattering is important.  
By varying an external magnetic field experimentalists can 
get a trapped Fermi gas with any scattering length.
({\it e.g.}\ \cite{Ohara:2002aaa}).

Depending on the scattering length, the ground state of cold 
Fermi gases can be very different.  
In two extreme cases, theoretical expansions in $|a|n^{1/3}\ll 1$ are
valid.  If there is a deep 2-fermion bound state ({\it i.e.}\ if
$0 < a n^{1/3} \ll 1$) then the fermion molecules form a 
Bose Einstein Condensate (BEC) at low temperatures. 
If there is no 2-body bound state ($-1 \ll an^{1/3} <0$) 
then delocalized Cooper pairs condense in a Bardeen-Cooper-Schrieffer
(BCS) superfluid.  What is the ground state for fermions when $|a|$
is large? No perturbative expansion converges for $|a|n^{1/3}>1$.
Ref.\ \cite{Chen:2003vy} is the only first
principles approach for theoretical calculation with $|a|n^{1/3} > 1$.
The case where $|a|\to \infty$ is especially interesting since 
systems should then exhibit universal behavior.

%\begin{figure}[th]
%\vspace{4.5cm}
%\special{psfile=OharaLithium.eps hscale=17 vscale=20 hoffset=0
%voffset=-20}
%\caption{\label{fig:OharaLithium} Scattering length between
%two ${}^6\mathrm{Li}$ hyperfine states as a function of external magnetic 
%field \cite{Ohara:2002aaa}.  Universal behavior is expected when the 
%scattering length diverges.}
%\end{figure}

%%%%%
%\vspace{-0.8cm}
\section{LATTICE THEORY}

Below we summarize the proposal of Chen and Kaplan \cite{Chen:2003vy}.
Consider 2 species of nonrelativistic fermions,
$\psi \equiv \left(\begin{array}{c}\psi_1\\ \psi_2\end{array}\right)$, 
with a local 4-fermion interaction with finite chemical potential $\mu$
\begin{eqnarray}
{\cal L} &=& \overline{\psi}_x\left(\psi_x - e^\mu\psi_{x-\hat{e}_0}\right)
\;-\; \frac{1}{2M}\,\overline{\psi}_x \left(\VEC{\nabla}^2\psi\right)_x
\nonumber \\
&+&\frac{C_0}{2}\left(\overline{\psi}_x e^\mu \psi_{x-\hat{e}_0}\right)^2 \, .
\end{eqnarray}
The 2 species can be envisioned as spin degrees-of-freedom, although
in the atomic gases they are two hyperfine states.   

In order to perform simulations
we must make the Lagrangian quadratic in the fermion field by introducing
an auxiliary real scalar field $\phi$ so that 
\begin{equation}
\frac{C_0}{2}\left(\overline{\psi}_x e^\mu \psi_{x-\hat{e}_0}\right)^2
\rightarrow \frac{1}{2}m^2\phi_x^2 \;-\; \phi_x\overline{\psi}_x e^\mu 
\psi_{x-\hat{e}_0}
\end{equation}
where we identify $m^2$ with $-C_0^{-1}$.

Now one introduces a source for the fermion pairing, a necessary step
to study spontaneous symmetry breaking in a finite volume
\begin{equation}
{\cal L}[J] ~=~ {\cal L} \;+\; \frac{1}{2}\left( J\,\psi^T\sigma_2 \psi
\;+\; J^*\,\overline{\psi}\,\sigma_2\, \overline{\psi}^T \right) \, .
\end{equation}
Combining $\psi$ and $\overline{\psi}$ into a 4-component Majorana
fermion, $\Psi^T \equiv (\psi^T, \overline{\psi}(i\sigma_2)^T)$,
the action can be written in terms of $4VN_t\times 4VN_t$ matrices
as $\Psi^T  {\cal A}  \Psi$, where
\begin{equation}
{\cal A} \equiv \left(\begin{array}{cc}
-i J & K^\dagger \\ K & -iJ^*\end{array}\right) \left(\begin{array}{cc}
i\sigma_2 & 0 \\ 0 & i\sigma_2\end{array}\right) \, .
\end{equation}
$K=K[\phi]$ represents the fermion kinetic energy and interaction terms,
and it is the only $2VN_t\times 2VN_t$ block to contain nontrivial spacetime 
terms.
The only coupling between species is through the $i\sigma_2$ blocks.
The partition function can be seen to have a real, nonnegative
integrand:
\begin{eqnarray}
{\cal Z}[J] &=& \int [d\phi][d\Psi] \exp\left(-\frac{1}{2}m^2\phi^2
- \frac{1}{2}\Psi^T{\cal A}\Psi\right) \nonumber \\
%&=& \int [d\phi] \;\mathrm{Pf}[{\cal A}] \;e^{-m^2\phi^2/2} \nonumber \\
&=& \int [d\phi] \;{\mathrm{det}}\left[|J|^2 + 
\tilde{K}^\dagger\tilde{K}\right] \;e^{-m^2\phi^2/2} 
\label{eq:Z}
\end{eqnarray}
where $K\equiv \tilde{K}\otimes I_2$ defines the $VN_t\times VN_t$ matrix
$\tilde{K}$.  We can perform Monte Carlo simulations of (\ref{eq:Z}) using
Hybrid Monte Carlo.

%%%%%
\section{PARAMETER SPACE}

We have 3 lattice parameters, $m^2$, $\mu$, and $M$ which control
the physical quantities $a$, $n$, and $T$ (ignoring finite $V$ and
setting $N_t$ fixed).  $1/a$ is matched to the
ratio $m^2/M$ at $T,\mu=0$ by requiring 2-body scattering amplitude
to match the leading term in the effective range expansion
as in pionless nuclear effective field theory \cite{Kaplan:1998we}.
$n$ is mainly controlled by tuning $\mu$.  In a
nonrelativistic theory, the fermion mass can be scaled away
by rescaling the time coordinate, so it can be interpreted as the
anisotropy.  For fixed $N_t$, increasing $M$ shortens the temporal
length, increasing $T$.

%%%%%
\section{CONTINUUM LIMIT}

\begin{figure}[t]
\vspace{6.7cm}
\includegraphics{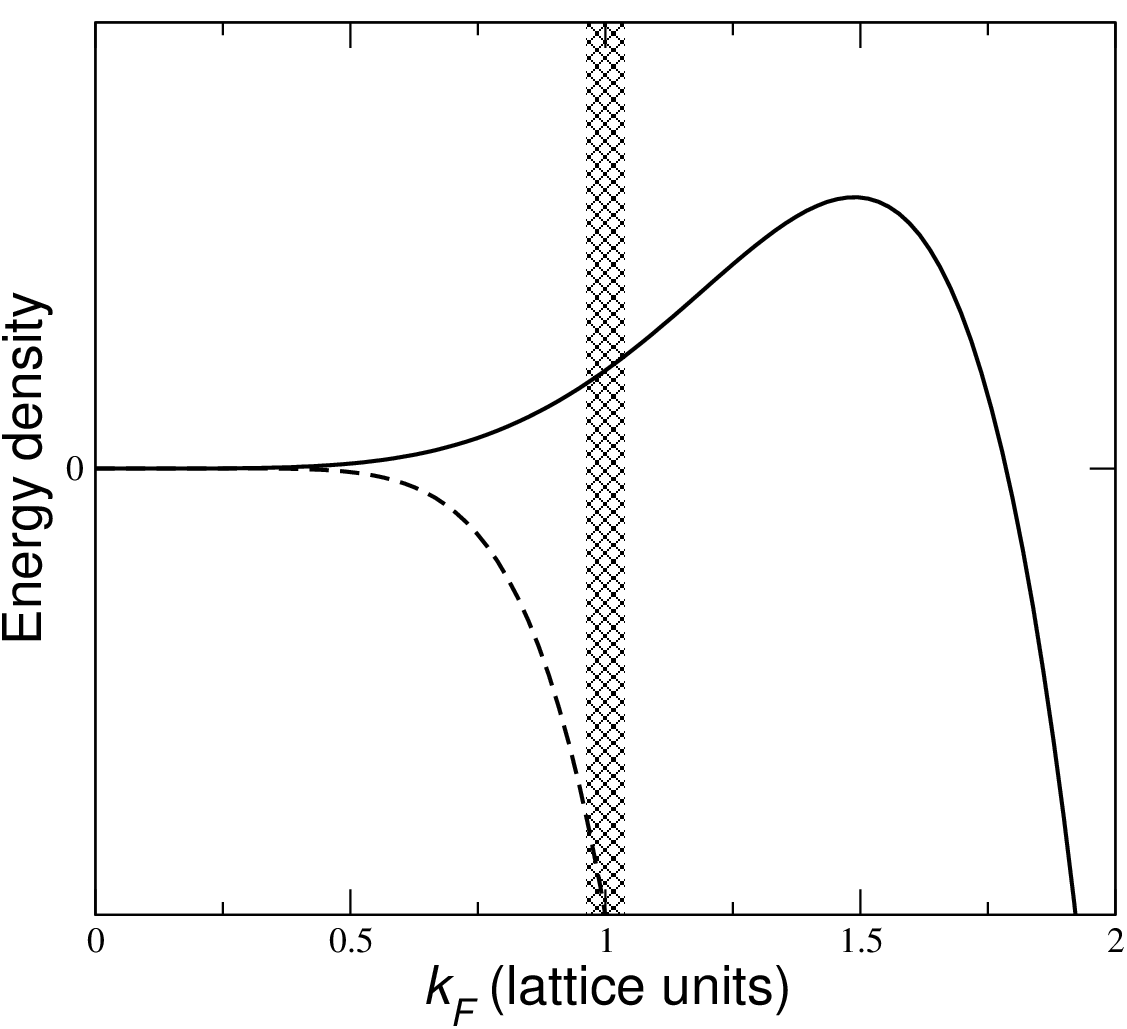}
\caption{\label{fig:MFinstability} Mean field illustration of the 
instability at strong coupling.  The solid line corresponds to a 
coupling which is small enough in lattice units for the ground state 
to be the physical one, the dashed line corresponds to a coupling so
strong that the ground state favors a fermion at every site even at
$\mu = 0$.}
\end{figure}

Some parts of parameter space do not describe the physics of a
dilute Fermi gas.
In a continuum theory with a local attractive interaction, the vacuum 
is unstable.  Since we are not interested in physics at short distances,
we simply have to regulate the theory.  
%For example, one might add a hard-core
%repulsive term to stabilize the ground state.  
In our case, the lattice
automatically provides a cutoff (see Fig.~\ref{fig:MFinstability}).

The continuum limit is defined as the limit where physical correlation
lengths $a, n^{-1/3}$, become infinitely large compared to the lattice 
spacing $b$.  This limit can only be taken using simulations
where one finds $n\,b^3 \ll 1$ at the same time that $1/a$ is tuned
to 0.  

\begin{figure}[t]
\vspace{6.7cm}
\includegraphics{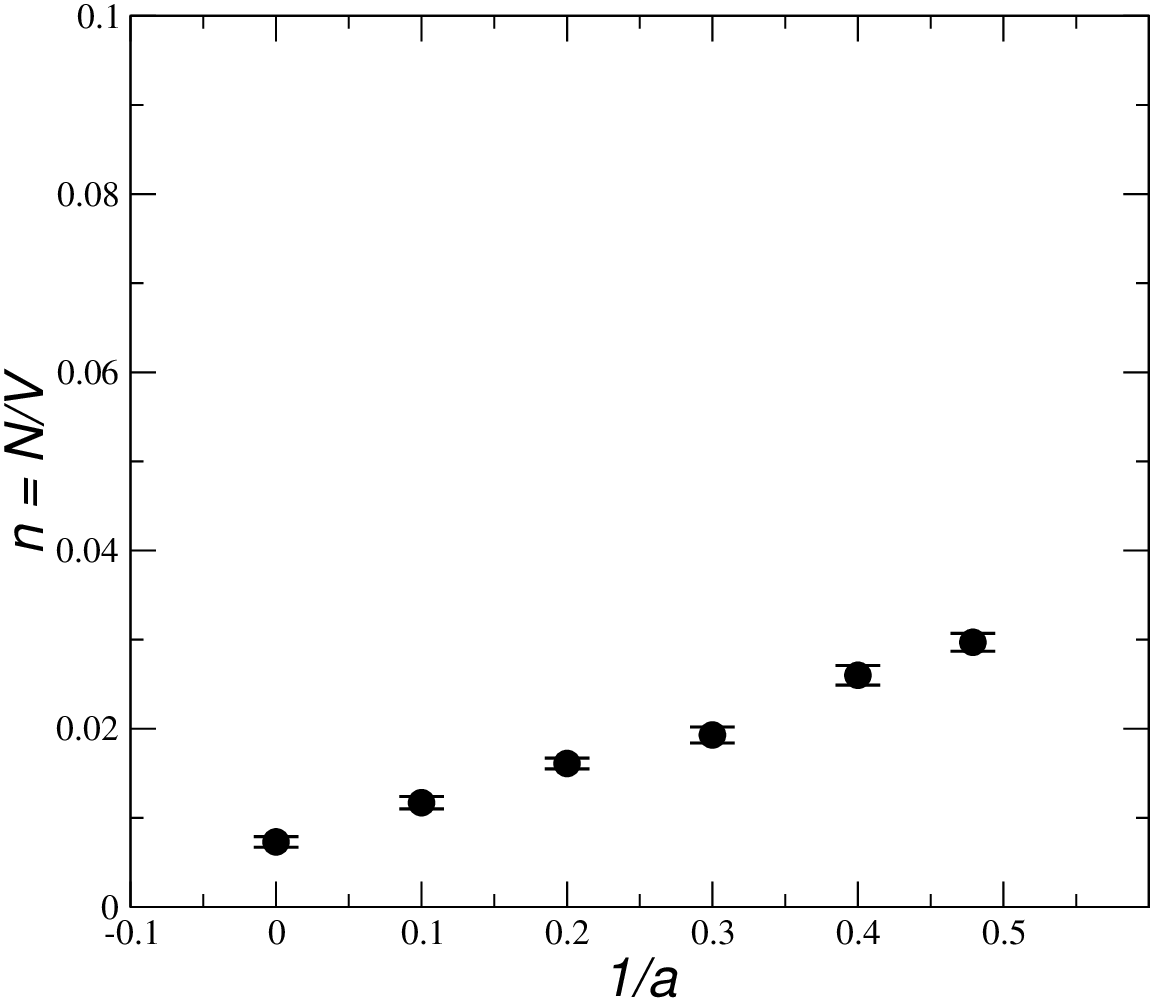}
\caption{\label{fig:density_ainv} Number density vs.\ inverse scattering
length (both in lattice units) computed on $6^3\times 12$ lattices
with $J=0.05$ and $\mu=0$.
Since $n\ll 1$ in lattice units, we can conclude the ground state is the
physical one and the continuum limit exists.
}
\end{figure}

Figure \ref{fig:density_ainv} shows
$n\ll 1$ (in lattice units)
at $1/a = 0$ (and $\mu = 0$) which is necessary and sufficient for
a continuum limit to exist.  $n$ is nonzero there due to finite
$V,N_t$ and $J$ effects.
As $1/a \to 1$, a rise in $n$ is observed.  It would be satisfying to
be able to increase $1/a$ enough so that we observe the artificial
first-order transition to the
filled-lattice state, but the HMC algorithm is 
becoming less efficient in this direction.  For larger
values of $1/a$ than plotted, 6000 HMC trajectories were not enough 
to equilibrate the lattice.
The breakdown of the HMC algorithm is known
from simulations of the Hubbard model.  Indeed the physics in 
the uninteresting (from our perspective) region of parameter space
is exactly that of the 3D attractive Hubbard model.  
 Having established the existence of a continuum limit
in the interesting part of parameter space, searching for the 
uninteresting region is not a priority.

%%%%%
\section{CRITICAL TEMPERATURE}

\begin{figure}[t]
\vspace{6.7cm}
\includegraphics{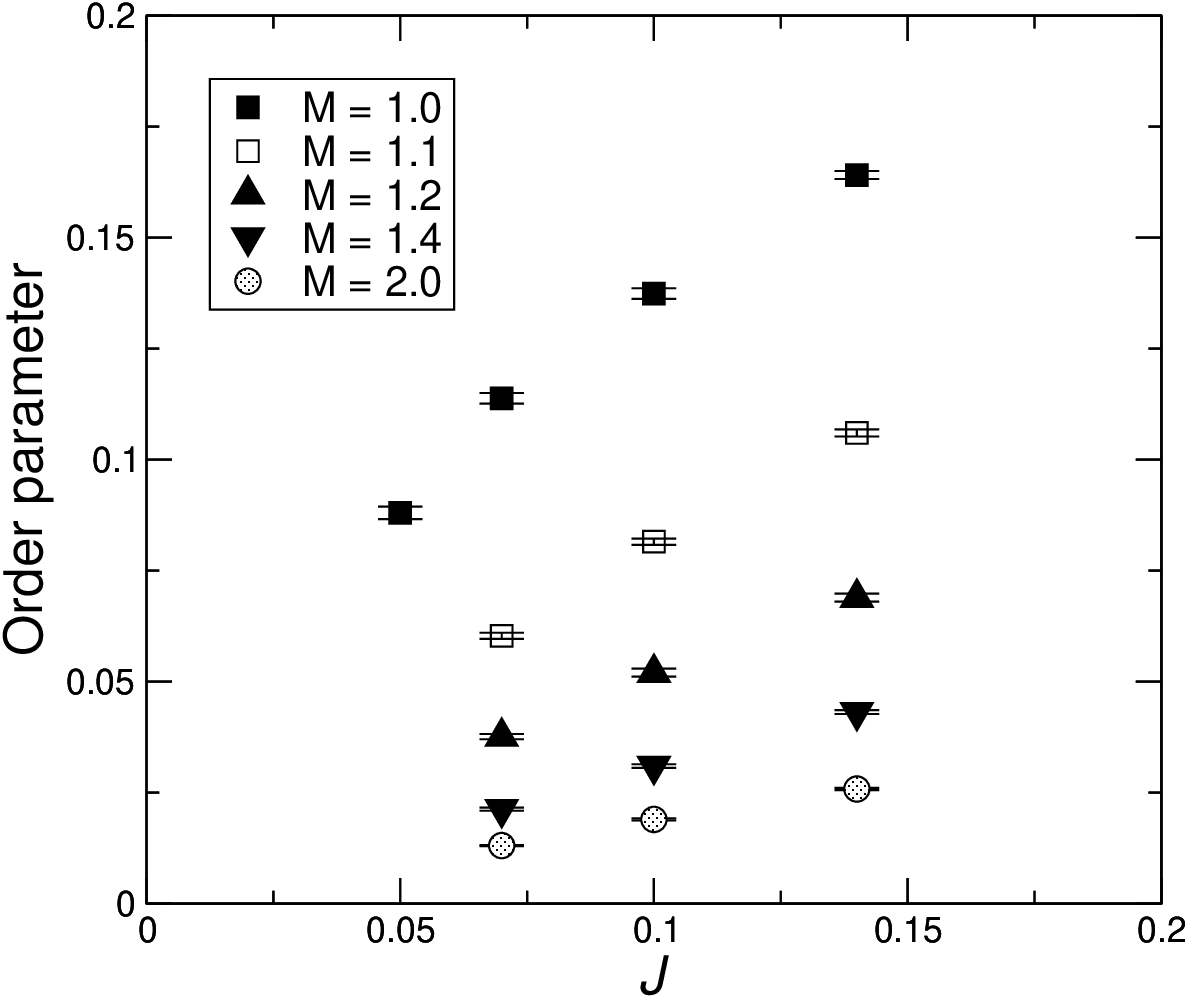}
\caption{\label{fig:order} Order parameter vs.\ pairing source $J$
(both in lattice units) computed on $8^3\times 16$ lattices
with $\mu=0.4$ and $m^2 = 0.1456$.  Increasing $M$ corresponds
to increasing $T$; linearly extrapolating the order parameter to
$J=0$ indicates U(1) restoration close to $M=1.4$.
}
\end{figure}

At low temperatures we expect condensation of fermion pairs, breaking
the global U(1) symmetry of fermion number conservation.  The order
parameter is 
\begin{equation}
\Sigma ~\equiv~ -\frac{1}{2}\left\langle \psi^T\sigma_2\psi 
\;+\; \overline{\psi} \,\sigma_2\,\overline{\psi}{\,}^T\right\rangle \,.
\end{equation}
Figure~\ref{fig:order} shows $\Sigma$ as a function of $J$ for several
values of $M$.  Extrapolating to $J=0$ linearly indicates
the vanishing of the order parameter as $M$ increases.  The validity
of the linear extrapolations is under study, since we expect finite
volume errors to increase as $J$ decreases \cite{Chen:2004aaa}.

%%%%%
\section{CONCLUSIONS}

The results of this exploratory study show
this lattice theory can describe the physics of dilute
Fermi gases.  These simulations show a continuum limit 
exists: it is defined by $b\,n^{1/3},b/a\to 0$ with fixed $a n^{1/3}$.
Study of the critical temperature has begun.  The first point
on the critical line has been located in the space of lattice
parameters $(m^2,M)$.  The critical line can then be mapped
to the physical space $(1/a,T)$.  For precision results,
one will have to repeat these steps for increasing volumes and
decreasing lattice spacings.  However, even with finite size and
spacing effects, understanding the behavior of $T_c$ as a function
of $an^{1/3}$ will be an important accomplishment.

%%%%%
\section*{ACKNOWLEDGMENTS}

I thank J-W.~Chen, Ph.~de~Forcrand, D.B.~Kaplan, and S.~Sharpe
for helpful conversations. Work was supported by 
DOE grant DE-FG02-00ER41132.

%%%%%%%%%%%%%%%%%%%%%%%%%%%


\begin{thebibliography}{20}

%\cite{Chen:2003vy}
\bibitem{Chen:2003vy}
J.-W.~Chen and D.B.~Kaplan,
%``A lattice theory for low energy fermions at finite chemical potential,''
Phys.\ Rev.\ Lett.\  {\bf 92}, 257002 (2004).
%[arXiv:hep-lat/0308016].
%%CITATION = HEP-LAT 0308016;%%

\bibitem{Ohara:2002aaa}
K.M. O'Hara {\it et al}., Phys.\ Rev.\ A {\bf 66}, 041401(R) (2002).

\bibitem{Kaplan:1998we}
D.B.~Kaplan, M.~Savage, and M.~Wise, Nucl.\ Phys.\ {\bf B534}, 329
(1998).

\bibitem{Chen:2004aaa}
J.-W.~Chen and M.~Wingate, in preparation.

\end{thebibliography}
\end{document}